\documentclass[preprint,11pt]{elsarticle}

\usepackage{amssymb}
\usepackage{amsmath,amsthm}
\usepackage[margin=1.4in]{geometry}

\journal{Transportation Research Part B: Methodology}

\begin{document}

\begin{frontmatter}

\title{Estimation of an Origin/Destination matrix: Application to a ferry transport data}

\author{Adrien Ickowicz, Ross Sparks}

\ead{adrien.ickowicz@csiro.au,ross.sparks@csiro.au}


\address{CSIRO Mathematics,\\
Locked Bag 17\\
North Ride, NSW\\
Australia}

\begin{abstract}
	The estimation of the number of passengers with the identical journey is a common problem for public transport authorities. This problem is also known as the Origin-Destination estimation (OD) problem and it has been widely studied for the past thirty years. However, the theory is missing when the observations are not limited to the passenger counts but also includes station surveys.\\
Our aim is to provide a solid framework for the estimation of an OD matrix when only a portion of the journey counts are observable.\\
Our method consists of a statistical estimation technique for OD matrix when we have the sum-of-row counts and survey-based observations. Our technique differs from the previous studies in that it does not need a prior OD matrix which can be hard to obtain. Instead, we model the passengers behaviour through the survey data, and use the diagonalization of the partial OD matrix to reduce the space parameter and derive a consistent global OD matrix estimator. 
	We demonstrate the robustness of our estimator and apply it to several examples showcasing the proposed models and approach. We highlight how other sources of data can be incorporated in the model such as explanatory variables, e.g. rainfall, indicator variables for major events, etc, and inference made in a principled, non-heuristic way.

\end{abstract}

\begin{keyword}
constraint maximum likelihood estimation \sep eigenvectors \sep counts estimation
\end{keyword}

\end{frontmatter}

\section{Introduction}

The Origin-Destination (OD) matrix is important in transportation analysis. The matrix contains information on the number of travellers that commute or the amount of freight shipped between different zones of a region. The OD matrix is difficult and often costly to obtain by direct measurements/interviews or surveys, but by using incomplete traffic counts and other available information one may obtain a reasonable estimate. A particular application of the OD matrix estimation is in the area of public transport. In order to improve their service, the responsible managers are looking for on-going evaluation of the passenger flow and the reasons that would influence this flow. This is typically the case for the City Rail, Sydney Bus and Sydney Ferry organisations, which handle the public transport in the region around the city of Sydney, Australia.

CityRail and Co are handling a large number of stations (wharfs, Bus stops) for Trains (Buses and Ferries) across the state. They carry thousands of passengers every day, and periodically optimise the time-table schedule to best meet the changing demand.\\
~\\
An ideal optimization of the schedule would consider the resources in trains, drivers, stations and passengers. While the primary informations (trains, drivers, stations) are known to CityRail and Co, the number of passenger on each train between each station cannot be deduced easily given their current passenger flow data collection processes.\\
~\\
Various approaches to estimating the OD matrix using traffic counts have been developed and tested \cite{Bera2011, Report} using traffic counts, or road traffic flows \cite{Dai2012}, \cite{Liu2011}. Most of the papers in the literature solve this problem by postulating a general model for the trip distribution, for example a gravity type model \cite{Chriqui1975,Tamin1989,Wilson2010,Carvalho2010}, which aims at introducing a prior knowledge on the traffic flows and assigning a cost to each journey. Then the inference is produced to estimate the parameters of this model. All these papers \emph{are not passengers oriented}.\\

Most of the work relating to OD matrix estimation are based on passengers observations assuming the knowledge of where the people get in and out of the public transport. 
Lo et al \cite{Lo1996} developed a framework centred on the passenger choice, which they called the random link choice, and model this to obtain a maximum likelihood estimator. Nandi et al \cite{Nandi2009} applied a strategy centred on a fixed cost per person per kilometre assumption on the air-route network of India and provide some comparisons with the real data.\\

When the information is not available (for example we have no data on when passengers get off the bus), Kostakos \cite{Kostakos} offers to use a wireless detection of the passengers' trips, and Lundgren and Peterson's model \cite{Lundgren2008} is based on a target OD-matrix previously defined. However, none of the cited work considered using survey data. Indeed, if no complete information is available about the passengers' destinations, the simplest solution is to use an appropriate survey to estimate destination information. Furthermore, what characteristics of the survey are required for the estimation to be accurate?

Bierliaire and Toint \cite{Bierlaire1995} introduces a structure-based estimation of the Origin-Destination matrix based on parking surveys. In their article, they used the parking surveys to infer an a priori estimate of the OD matrix, and they used this prior in coordination with the partial observations of the network flows to derive a generalized least square estimator of the OD matrix. Despite its novelty, this article assume that the behaviour of car-user and public transport users are the same, at least regarding their respective OD matrix. Given that the public transport network topology is often different from the road network topology, one may doubt the accuracy of this assumption. Moreover, they just use the partial structure extracted from the surveys.\\

The purpose of this paper is then to develop an estimation procedure for the Origin-Destination matrix based on the ticket records available for the transport network and/or on previous surveys. Unlike the article from Bierliaire \cite{Bierlaire1995}, we use survey data collection from public transport users, and estimate the approximate whole matrix structure through the estimation of its eigenvectors. We propose a robust version of the estimator to avoid biases induced by the survey. We also construct a regression estimation procedure that accounts for the influence of exogenous variable such as the weather conditions or the time of the year. 
~\\
We first briefly present the passenger model, and then move on to outlining the observations model. In section \ref{sec:OM}, we explain how the measurements are obtained, and what measurements error should be expected.  In section \ref{sec:mam}, we explain the assumptions we make on the measurements, and how this affects our estimation procedure. We present in section \ref{sec:est} the maximum likelihood (ML) estimation procedure, by providing a system of equation to be solved, for deriving estimators. We improve on this ML estimation to make it robust to survey biases in section \ref{sec:rob}.  Finally, we present  a simulation example and an application to a real world case in section \ref{sec:app}. We finally comment on the results and outline some future research opportunities.

\section{The Passenger model}
\label{sec:PM}
Let $\mathcal{M}^t$ be the matrix of passengers number between the stations in the rail network over time period $t$ so that ${\bf m}_{ij}^t$ is the number of passengers who depart from station $i$ and arrive at station $j$ at time period $t$. Given that there is an obvious time dependency here, denoted by $t$ the period in which the commuting occur (for example a day). The purpose of this work is to provide an estimation of $\mathcal{M}^t$ given the observations specified in section \ref{sec:OM}.

\section{The observations model}
\label{sec:OM}

The observations provided about the passengers are very different, and only considering them all allow a direct estimation of $\mathcal{M}^t$. We list in the subsections \ref{OM-casual},\ref{OM-deparr} and \ref{OM-regular} the different kind of observations.

\subsection{Casual commuters}
\label{OM-casual}

A casual commuter is defined as a single or return journey that is not repeated regularly (e.g. daily). Typically, people going to a once-in-a-year event will buy their ticket for that trajectory and will probably return on the same day. Accordingly for single and day return tickets, we have complete information under the assumption that they take the next train after purchasing their ticket and that they take the shortest route.
Let $\mathcal{X}_C^t$ be that matrix of measurements.

\subsection{Departure-Arrival recordings}
\label{OM-deparr}

Each journey between major stations, the passenger has to validate his ticket through the machines at the entrance of the station, and do it again at the exit. Between minor stations we assume they take the next train to arrive at the station they purchased their ticket at and assume they take the trio planners recommended route for that time. Two scenarios are considered. In the first one, (called $\mathcal{H}_a$), every station in the network have these machines. In the second case (called $\mathcal{H}_b$) only major stations have these machines. In any case, let call $\mathcal{Y}_D^t$ the vector corresponding to the departures at the stations, and $\mathcal{Y}_A^t$ the vector of arrivals. 

\subsection{Regular commuting partial measurements}
\label{OM-regular}
Fortunately we can have regular passengers with specific departure and destination, and this matrix will be denoted $\mathcal{X}_{Rs}^t$, where the rows stand for the departure stations and the columns for the arrival stations. This matrix is observed, and assumed distributed according to a Poisson probability function with mean $Rs$.\\
The main part of the information, however, remains unknown. Indeed most of the passengers will probably have a Zone ticket for a period of time, from 1 week to 1 year. The nature of these tickets make the station of departure and arrival unknown, and is the main challenge of this paper. Let call $\mathcal{X}_{Rz}^t$ the matrix of Zone passengers numbers.\\
To make a proper statistical inference, we need two assumptions;
\begin{itemize}
\item The traveller will act independently of the validity duration of his ticket;
\item The regular traveller commits to a return journey on each working day.
\end{itemize}
The observations linked to this model are two-folds. For major stations, we have the total number of passengers that crossed the boom gates, in and out. For stations without boom gates, the observations have to estimated using a survey. We also have access to the total number of people with a valid zone ticket at time $t$ (e.g. the day of the analysis), denoted $N_{Rz}^t$,
\begin{eqnarray}
N_{Rz}^t &=& \mathbf{1}' \mathcal{X}_{Rz}^t \mathbf{1}.
\end{eqnarray}
In the end, the total number of regular passenger at the time period $t$ will be denoted $\mathcal{X}_{R}^t$, and we have,
\begin{eqnarray}
\mathcal{X}_{R}^t &=& \mathcal{X}_{Rz}^t + \mathcal{X}_{Rs}^t.
\end{eqnarray}

\section{Model assumptions and modelling}
\label{sec:mam}

With these very different observations, we need a good fitting model based on reasonable assumptions. Sections \ref{MAM-gm}, \ref{MAM-cm}, \ref{MAM-da} and \ref{MAM-rm} presents these assumptions for each parameter in our model.

\subsection{General Modelling Assumptions}
\label{MAM-gm}
Recall that $\mathcal{X}_{M}^t$ is a matrix of count, the main assumption on that matrix is that the number of passenger is the sum of the casual passengers ($\mathcal{X}_{C}^t$) plus the regular passengers ($\mathcal{X}_{R}^t$) plus a matrix stating the unusual big events such as major sporting events, or large concerts (called $\mathcal{X}_{E}^t$),

\begin{eqnarray}
\mathcal{X}_{M}^t &=& \mathcal{X}_{C}^t + \mathcal{X}_{R}^t + \mathcal{X}_{E}^t.
\end{eqnarray}

\subsection{Casual Commuter Matrix Assumptions}
\label{MAM-cm}

The casual commuter journey could be assumed to be Poisson distributed i.e. $\mathcal{X}_C \sim \mathcal{P}(C)$\footnote{This is a notation abuse stating that every element in $\mathcal{X}_C$ is supposed to be drawn with a Poisson distribution which parameter belong to the matrix $C$.} where $C$ is the matrix of means for the counts. \\
However, the variance of the counts are not expected to be equal to their mean and so the Poisson counts assumption may be unrealistic. Therefore, we decided to use a Negative Binomial regression model for $\mathcal{X}_C$, which can be over-dispersed in order to better describe the distribution of the counts. We specify that $C$ is distributed according to a Gamma distribution, $C \sim \mathcal{G}a(r_C,p_C/(1-p_C))$. For a purpose of simplicity, let $\mathcal{X}_C$ be distributed as negative binomial with parameters $r_C$ and $p_C$ (we will denote $\mathcal{X}_C \sim \mathcal{NB}(r_C,p_C)$).

\subsection{Departure-Arrival Assumptions}
\label{MAM-da}

According to the definition of the measurements, the following relationships hold:
\begin{equation}
\label{eq:ODeq}
\left\{
\begin{array}{ll}
\mathcal{X}_R^t \mathbf{1} = \mathcal{Y}_D^t \\
\mathbf{1}' \mathcal{X}_R^t = \mathcal{Y}_A^t
\end{array}
\right.
\end{equation}
where $\mathcal{Y}_D^t$ and $\mathcal{Y}_A^t$ are the vectors of the total number of departures and arrivals at each station during time period $t$.

\subsection{Regular Commuter Matrix Assumptions}
\label{MAM-rm}

For the same reasons as described for the casual matrix, we will use a negative binomial distribution to model the uncertainty around the regular traveller's information. However, unlike the casual commuter, we do not suspect an over-dispersion but an under-dispersion, so that, $\mathcal{X}_{Rs} \sim \mathcal{NB}(r_{Rs},p_{Rs})$ and $\mathcal{X}_{Rz} \sim \mathcal{NB}(r_{Rz},p_{Rz})$. Let $R_s$ and $R_z$ be the expectation of $\mathcal{X}_{R_s}$ and $\mathcal{X}_{R_z}$.

\section{Naive Maximum Likelihood (ML) Estimation}
\label{sec:est}
When the model is well defined, the estimation procedure is computationally straight forward, e.g., between major stations where we have complete information of arrivals and departures. Meaning that the maximum likelihood estimation method accuracy, practically depends on the efficient solving of the optimization problem. In this section, the stationary model parameters are estimated from the data. Since the process is unlikely to be stationary, we present a second option (Section \ref{Est:reg}), a multivariate spatio-temporal model that we expect to fit the data better.  

The estimation procedure will be carried out in well-defined steps. If we ignore the time dependence, the successive observations can be considered independent, identical random counts from Negative Binomial or Poisson distribution. This means that simple maximum likelihood estimation should work well, especially for large sample sizes.

\subsection{Casual parameters estimation}
\label{sec:CPE}
We observe $\mathcal{X}_{C}^t$ for several realizations. Given no space-time dependencies we assume that $\mathcal{X}_C^t$ is independently distributed as $\mathcal{NB}(r_{C},p_{C})$. The likelihood is then,
\begin{eqnarray}
L\big((x_C^t)_{\{t_1,...,t_n\}} \vert r_C, p_C \big) & = & \prod_t f_C(x_C^t)\\
{} &=& \prod_t \Big[ \frac{\Gamma(r_C+ x_C^t)}{\Gamma(r_C) x_C^t!} p_C^{r_C} (1-p_C)^{x_C^t} \Big]
\end{eqnarray}
where $x_C^t$ stands for one element of the matrix $\mathcal{X}_C^t$. We thus can estimate the parameters through,
\begin{eqnarray}
(\hat{r}_{C}, \hat{p}_{C}) &=& \underset{r,p}{\operatorname{argmax}} \quad L\big((x_C^t)_{\{t_1,...,t_n\}} \vert r_C, p_C \big)
\end{eqnarray}
Despite the absence of closed form solution to this problem, the optimization algorithms can quickly lead to a global maximum.

\subsection{Regular parameters estimation}

Unfortunately we don't have complete information for those with weekly,  monthly , quarterly or anuual tickets (long-term tickets).  We have information of the times they enter and departs at major stations but we don't have complete information for the long term tickets either to or from minor stations.  Our assumption here is that only a proportion $\pi_z$ of the $R_z$ people will travel on day $t$, where $0 \leq \pi_z \leq 1$. $\pi_z$ is an additional parameter that reflects the passengers habit. It does exist because when performing the estimation, one may find a bigger estimation of travellers than what is observed. Some of the difference is due to the randomness of $\mathcal{X}_{R}$, but it might also be explained by the fact that travellers with prepaid long term tickets will not necessary travel each of the working day of the week. \\
However, we may provide the same estimation for the $R_s$ parameters as we did in the previous section, that is,
\begin{eqnarray}
(\hat{r}_{Rs}, \hat{p}_{Rs}) &=& \underset{r,p}{\operatorname{argmax}} \quad L\big((x_{Rs}^t)_{\{t_1,...,t_n\}} \vert r_{Rs}, p_{Rs} \big)
\end{eqnarray}
where $L$ stands for the same likelihood function as above.\\

This leads us to the final estimation, the contribution this paper makes to the literature. The aim is to estimate the matrix $\mathcal{X}_{Rz}^t$ with the available departure and arrival data. The first step is to estimate the general shape of the $\mathcal{X}_{Rs}^t$ matrix. The problem is to achieve this in a simple way given that $\mathcal{X}_{Rz}^t$ is to be estimated with $N \times N$ parameters, and only $N$ equations. The following paragraph presents an elegant solution to this problem. \\
Recall $R_z$ as the expectation of $\mathcal{X}_{R_z}^t$. It is assumed symmetric, we can diagonalize it, so that,
\begin{eqnarray}
\label{eq:diag}
R_z &=& P D_z P^{-1}
\end{eqnarray}
where $P$ is a projection matrix of eigenvectors of $R_z$ and $D_z$ is a $N \times N$ diagonal matrix, with terms equal to the respective eigenvalues. Therefore, if the structure of $R_z$ is known (i.e. the eigenvectors are known) and constant, then we have reduced the problem to solving a system of $N$ unknown parameters with $N$ equations . Given Eq. \ref{eq:ODeq} and the previous estimations, we have the following system,
\begin{equation}
\label{eq:ODeq2}
\left\{
\begin{array}{ll}
\mathcal{X}_{Rz}^t \mathbf{1} = \tilde{\mathcal{Y}}_D^t \\
\mathbf{1}' \mathcal{X}_{Rz}^t = \tilde{\mathcal{Y}}_A^t
\end{array}
\right.
\end{equation}
where $\tilde{\mathcal{Y}}_D^t$ and $\tilde{\mathcal{Y}}_A^t$ are obtained by simple subtraction. The probability density function of the observations $\tilde{\mathcal{Y}}_D^t,\tilde{\mathcal{Y}}_A^t$ can then be written,
\begin{eqnarray}
\forall i\in [1,n] \quad p \big( y_{Di}^t \vert R_z, p_{Rz} \big) &\sim & \mathcal{NB}( \sum_j r^{ij}_z, p_{Rz})
\end{eqnarray}
where $(r^{ij}_z)_{(i,j)} = R_z$ and $(y_{Di}^t)_i = \tilde{\mathcal{Y}}_D^t$. According to this equation, we then have $n$ likelihood equations ($i \in [1,n]$) but $n(n-1)/2$ parameters to estimate ($R_z$ being symmetric). To perform the estimation, we then need to reduce the number of parameters. According to Eq. \ref{eq:diag}, we have,
\begin{eqnarray}
\label{eq:density}
\sum_j r^{ij}_z &=& \sum_{k} \lambda_k p_{ik} \sum_j p_{jk}
\end{eqnarray}
Then, if we knew $P$, the maximum likelihood would be tractable and provide an estimation of $R_z$ (with $\lambda_{i}$ being the $i$th eigenvalue). The $n$ likelihoods would look like, 
\begin{eqnarray}
\label{eq:like}
\forall i\in [1,n] \quad L \big( y_{Di}^t \vert R_z, p_{Rz} \big) & = & \prod_t p \big( y_{Di}^t \vert R_z, p_{Rz} \big)
\end{eqnarray}
where $p \big( y_{Di}^t \vert R_z, p_{Rz} \big) \sim \mathcal{NB}( \sum_{k} \lambda_k p_{ik} \sum_j p_{jk} , p_{Rz})$. The maximum likelihood estimating equations are then,
\begin{equation}
\label{eq:mle0}
(\hat{\lambda}_{1}, \dots ,\hat{\lambda}_{N}, p_{Rz}) = \underset{\lambda_i, p_{Rz}}{\operatorname{argmax}}
\left|
\begin{array}{l}
L \big( y_{D_1}^t \vert R_z, p_{Rz} \big)\\
\vdots \\
L \big( y_{D_n}^t \vert R_z, p_{Rz} \big)
\end{array}
\right.
\end{equation}
Moreover, this complex equation can be simplified by assuming that the observations are independent conditionally on knowing the parameters (e.g. the errors are independents), and then,
\begin{eqnarray}
\label{eq:mle}
(\hat{\lambda}_{1}, \dots ,\hat{\lambda}_{N}, p_{Rz}) &=& \underset{\lambda_i, p_{Rz}}{\operatorname{argmax}} \prod_i L \big( y_{D_i}^t \vert R_z, p_{Rz} \big)
\end{eqnarray}
The dimension of the parameter space is reduced according to the knowledge of $P$. To estimate $P$, remember that $R_s$ is symmetric, we have,
\begin{eqnarray}
\label{eq:diag2}
R_s &=& \tilde{P} D_s \tilde{P}^{-1}
\end{eqnarray}
Then, we make the assumption that all the regular passengers behave identically over time, that is, $P$ is not a function of time. Then we have, 
\begin{eqnarray}
\label{eq:ass1}
P &=& \tilde{P} 
\end{eqnarray}
Then $P$ can be estimated from $\hat{R}_s$, and introduced into Equation \ref{eq:density} to obtain Eq. \ref{eq:mle}. Finally, we can use $\hat{D}_z$ and $\hat{P}$ to estimate $\hat{Rz}$. 

\subsection{Optimization Issues}
\label{sec:OI}
Finding the solution of equation \ref{eq:mle} is a classic optimization problem. The simple likelihood shape insures the existence of a solution, and it can be found by any standard optimization function. Let this solution be $\hat{Rz}^s$. The problem with this estimator is that we can not guarantee that it will fulfil the underlying constraint of the density parameters. Indeed, some values in the matrix will be negative, and there will be some element in the diagonal that won't be null. Therefore, some constraints have to be added to the ml estimating equations. These are:\\
{\bf Constraint 1}
\emph{All the elements in the matrix $\hat{R}z$ are greater than or equal to zero, or equivalently,
\begin{eqnarray}
\forall (i,j) \quad \sum_{k} \lambda_k p_{jk} p_{ik} & \geq & 0 
\end{eqnarray}
}\\
{\bf Constraint 2}
\emph{All the diagonal elements in the matrix $\hat{Rz}$ must be zero, or equivalently,
\begin{eqnarray}
\forall i \quad \sum_{k} \lambda_k p_{ik}^2 & = & 0 
\end{eqnarray}
}
\\
{\bf Constraint 3}
\emph{The last set of parameters to be estimated is the probability matrix $p_{R}$. Therefore, all the elements should belong to the interval $[0,1]$,
\begin{eqnarray}
\forall (i,j) \quad p_R^{ij} & \in & [0,1] 
\end{eqnarray}
}
Most of the optimization algorithms that deal with the constraint require an initialization which belong to the constrained space. One could be tempted to address as a starting point the mean value of the observations, according to the one-dimensional ($n=1$) result. However, it is very unlikely that this initial point will satisfy the constraints. Therefore, the best choice so far seems to be the diagonal elements of the matrix $D_s$, given that they naturally fill {\bf Constraint 1} and {\bf Constraint 2}.\\
~\\
The complete optimization program therefore becomes,
\begin{equation*}
\left |
\begin{aligned}
& \underset{\Lambda}{\text{maximize}}
& &  \prod_i L \big( y_{D_i}^t \vert R_z, p_{Rz} \big)\\
& \text{subject to}
& & {\bf C1},{\bf C2},{\bf C3}
\end{aligned}
\right.
\end{equation*}
with the initial value $\Lambda_0 = diag(D_s)$. This optimization program can be replaced by an explicit expression of the estimator, subject to some constraints stated in \ref{ann:A}. The main constraint is the Poisson distribution assumption, so that we have,\\

{\bf Proposition 1:} \emph{Assume that $g \sim \mathcal{P}()$, then
\begin{eqnarray}
\label{eq:linsol_lamf}
\hat{\Lambda} & = & S_d^{-1} {}^t P [ diag(PS) ]^{-1} \bar{Y}
\end{eqnarray}
\qquad where $\hat{\Lambda}$ is the matrix of estimated eigenvalues of $R_z$.}\\

If now we consider a Gaussian likelihood instead of Poisson, the following maximum likelihood estimator is found,\\

{\bf Proposition 2:} \emph{Assume that $g \sim \mathcal{N}()$, then
\begin{eqnarray}
\label{eq:linsol_lamfg}
\hat{\Lambda}_{\mathcal{N}} & = & S_d^{-1} {}^t P \bar{Y}
\end{eqnarray}
\qquad where $\hat{\Lambda}_{\mathcal{N}}$ is the matrix of estimated eigenvalues of $R_z$.}\\

The proofs of Propositions 1 and 2 are presented in \ref{ann:A}. We can also derive the follwing theorem, that ensures us of the quality of the estimation,\\

{\bf Theorem 1}\\
Assume that $\hat{P} \xrightarrow[]{\ a.s.\ } P$ (see Anderson \cite{Anderson1963},\cite{Anderson1987}). Then we have,
\begin{eqnarray}
\hat{\Lambda}_{\mathcal{N}} \xrightarrow[]{\ \mathcal{P} \ } \Lambda
\end{eqnarray}
The proof is presented in \ref{ann:B}.

\subsection{Multivariate estimation}
\label{Est:reg}

If we want to deal with a more realistic modelling, it seems obvious that we have to consider spatial, temporal and multivariate influences on the number of passengers, and then on the parameters of our modelling. \\
Let $(X_i)_i$ be a set of exogenous features, and $(\beta_i)_i$ their corresponding (unknown) influence on the number of passengers. Therefore, the regression model can be written,\\
\begin{eqnarray}
\forall i\in [1,n] \quad p \big( y_{Di}^t \vert R_z, p_{Rz}, (X_i)_i \big) &\sim & \mathcal{NB} \big( \sum_{k} \lambda_k p_{ik} \sum_j p_{jk} + \sum_l X_l \beta_l, p_{Rz}\big)
\end{eqnarray}
where the parameters to be estimated are $\theta = \big\{ (\lambda_k)_k, (\beta_l)_l, p_{Rz} \big\}$. The likelihood is expressed as in Eq. \ref{eq:like}, and we have,
\begin{eqnarray}
\label{eq:mle_multi}
((\hat{\lambda}_{i})_i ,(\hat{\beta}_{l})_l,p_{Rz}) &=& \underset{\theta}{\operatorname{argmax }}\quad \prod_i L \big( y_{D_i}^t \vert R_z, p_{Rz}, (X_i)_i \big)
\end{eqnarray}
Then, if we consider the projection matrix to be constant over time, then we have the following proposition (proved in \ref{ann:C}),\\

{\bf Proposition 3:}
\emph{If we apply the same methodology as before we obtain the following estimator,
\begin{eqnarray}
\hat{\Lambda}_{R} & = &  S_d^{-1} {}^t P  [diag(P S)]^{-1} Y {}^t X (X {}^t X)^{-1}
\end{eqnarray}
where $X = (x_m^l)_{(m,l)}$ is the matrix of the exogenous variables.}\\

A more complex solution would be to perform the same technique as before, except that $P$ will no longer be constant. Therefore, the varying $P_t$ estimated from $R_s$ will help in the recovery of the matrix $R_z$. The reason why we cannot perform a direct estimation of the $\beta$'s through this technique is because $R_z$ can be diagonalize, but we don't know how the $\beta$ influence the $\lambda$. Then a space-time model for the regular zone passenger is required. A further sensitivity analysis has to be run in order to figure out if the results are significantly affected by the increased uncertainty.

\section{Ad hoc Approach}
\label{sec:rob}
The quality of the estimator will strongly depend on the survey sub-matrix $R_s$. In section \ref{sec:est}, we assume that the similarity between the matrix $R_s$ and $R_z$ relies on the projection matrix $P$ and $\tilde{P}$. Moreover, we assume $P = \tilde{P}$ (Eq. \ref{eq:ass1}), and this assumption is the key to derive the Propositions $\bf 1-3$.\\
It is however difficult to design and implement a survey that provides accurate information. Therefore, Eq. \ref{eq:ass1} is no longer valid. To overcome this, we propose to consider an Ad hoc estimation of the O-D matrix.\\

\subsection{Estimator}

Despite the survey having some unknown biases, it provides useful information that we need. Let $\hat{\Lambda}_{\mathcal{N}}$ be the estimation based on the eigenvalues of $R_z$, relying on the assumption $P = \tilde{P}$. While inaccurate, this estimation looks like the perfect prior information. We define,
\begin{eqnarray}
R_z^{\pi_0} &=& \tilde{P} \times \textrm{diag}(\hat{\Lambda}_{\mathcal{N}}) \times \tilde{P}^{-1},
\end{eqnarray}
Now, given that the survey may be biased, we need to emphasize the influence of the observations $\mathcal{Y}$, by building an observation-based matrix $\hat{R}_z^{ob}$, such that,
\begin{eqnarray}
\hat{r}_{ij}^{ob} &=& \frac{1}{2n} \Big( \bar{s}_{i.} + \bar{s}_{.j} \Big),
\end{eqnarray}
where $\bar{s}_{i.} = 1/ n_{obs} \sum_k \sum_j x_{ij}^k$ and $\bar{s}_{.j} = 1/ n_{obs} \sum_k \sum_i x_{ij}^k$. $R_z^{\pi_0}$ is symmetric, and roughly correspond to an equal partition of the passengers in the different stations. Let $(\hat{r}^{\pi_0}_{ij})_{ij} = R_z^{\pi_0}$. Now, we need to integrate the information provided by $R_z^{\pi_0}$. Then we define the final ad hoc estimator $\hat{R}_z^{ah}$,
\begin{eqnarray}
\hat{r}_{ij}^{ah} &=& \frac{n}{2} \Big( \frac{\hat{r}_{ij}^{ob} \times \hat{r}^{\pi_0}_{ij}}{\sum_j \hat{r}_{ij}^{ob}} + \frac{\hat{r}_{ji}^{ob} \times \hat{r}^{\pi_0}_{ji}}{\sum_j \hat{r}_{ji}^{ob}} \Big).
\end{eqnarray}

\subsection{Robustness Analysis}

\begin{figure}[htb]
\centering
\includegraphics[width=7cm, height=6cm]{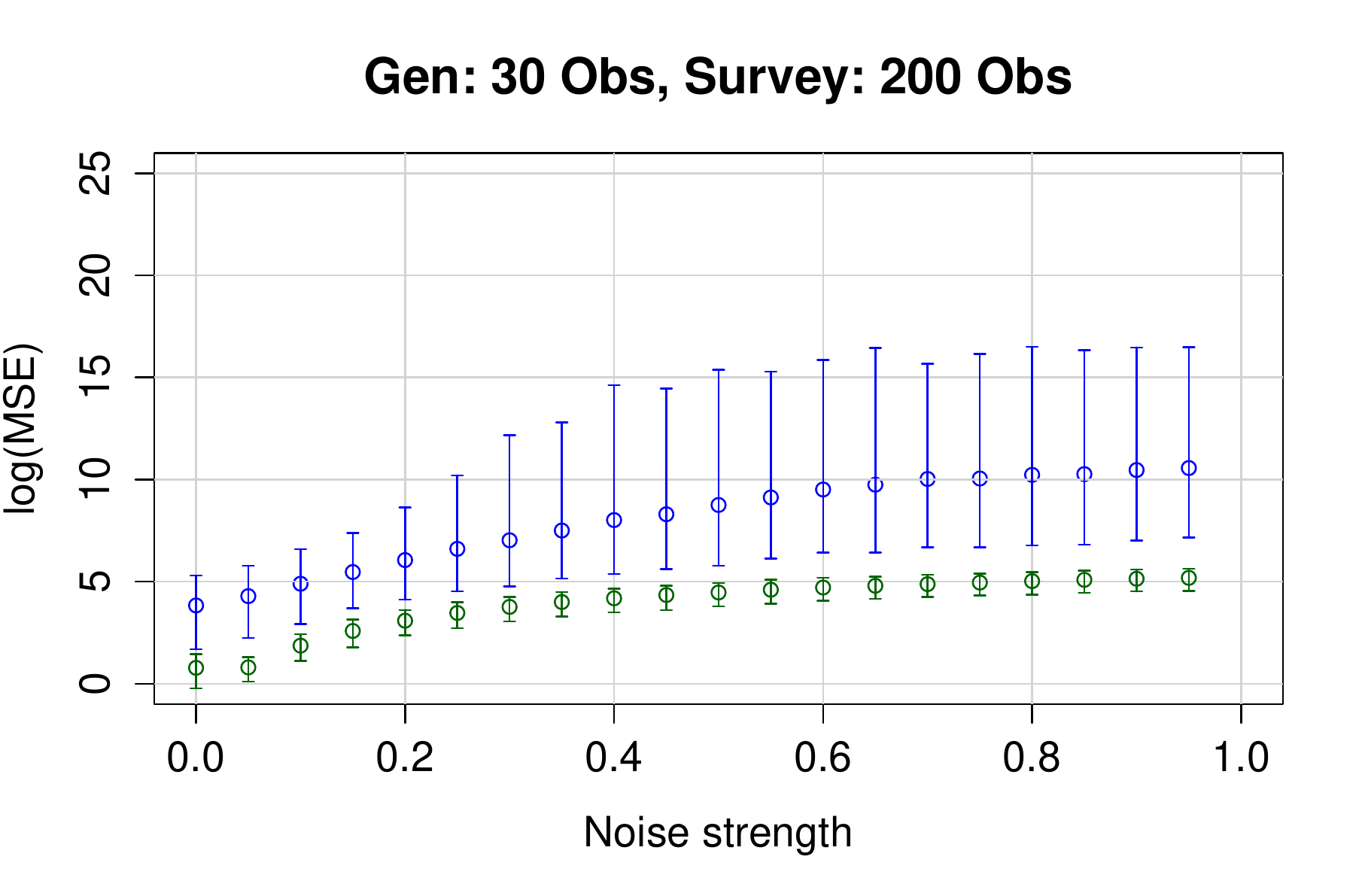}
\includegraphics[width=7cm, height=6cm]{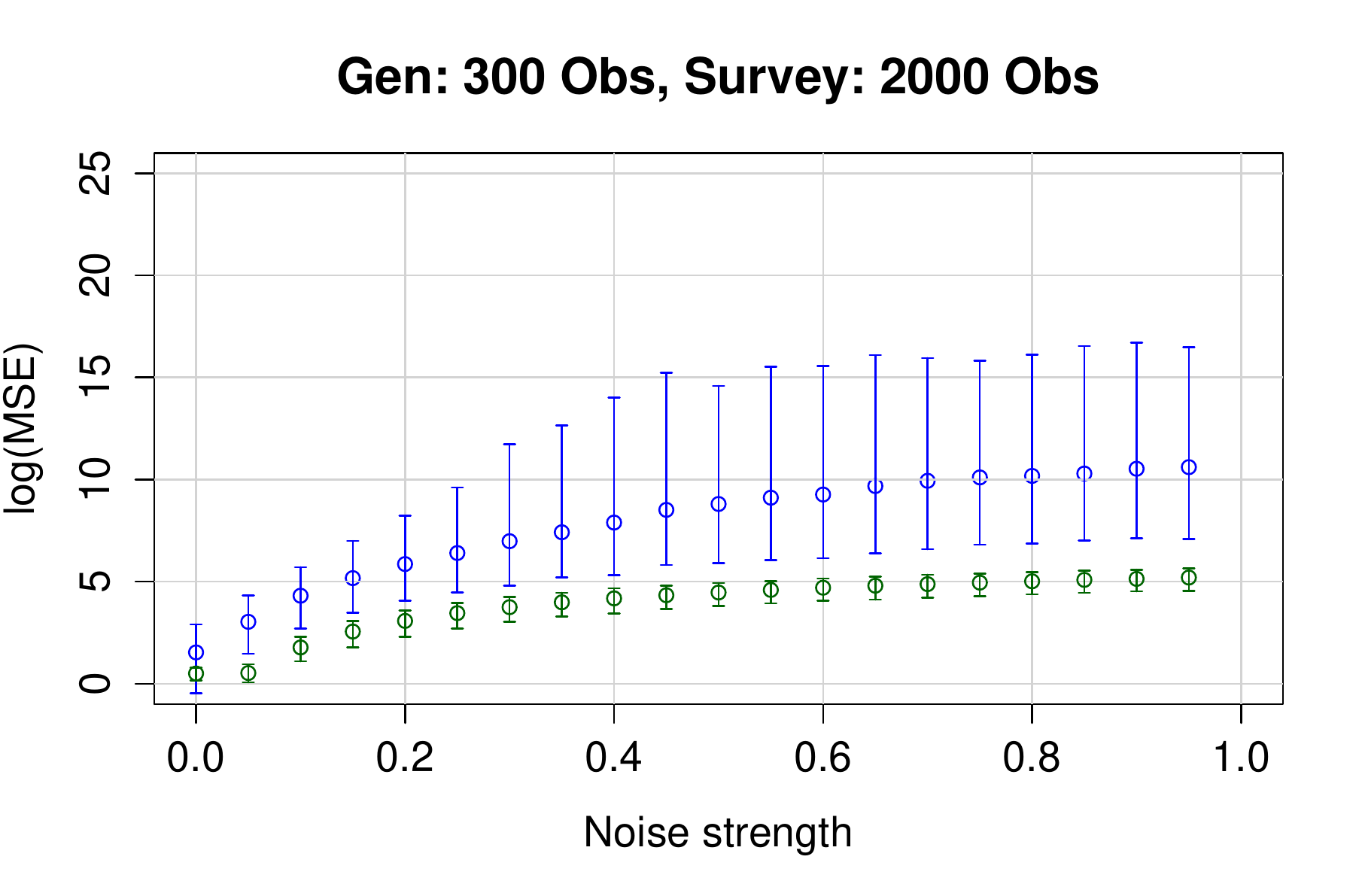}\\
\includegraphics[width=7cm, height=6cm]{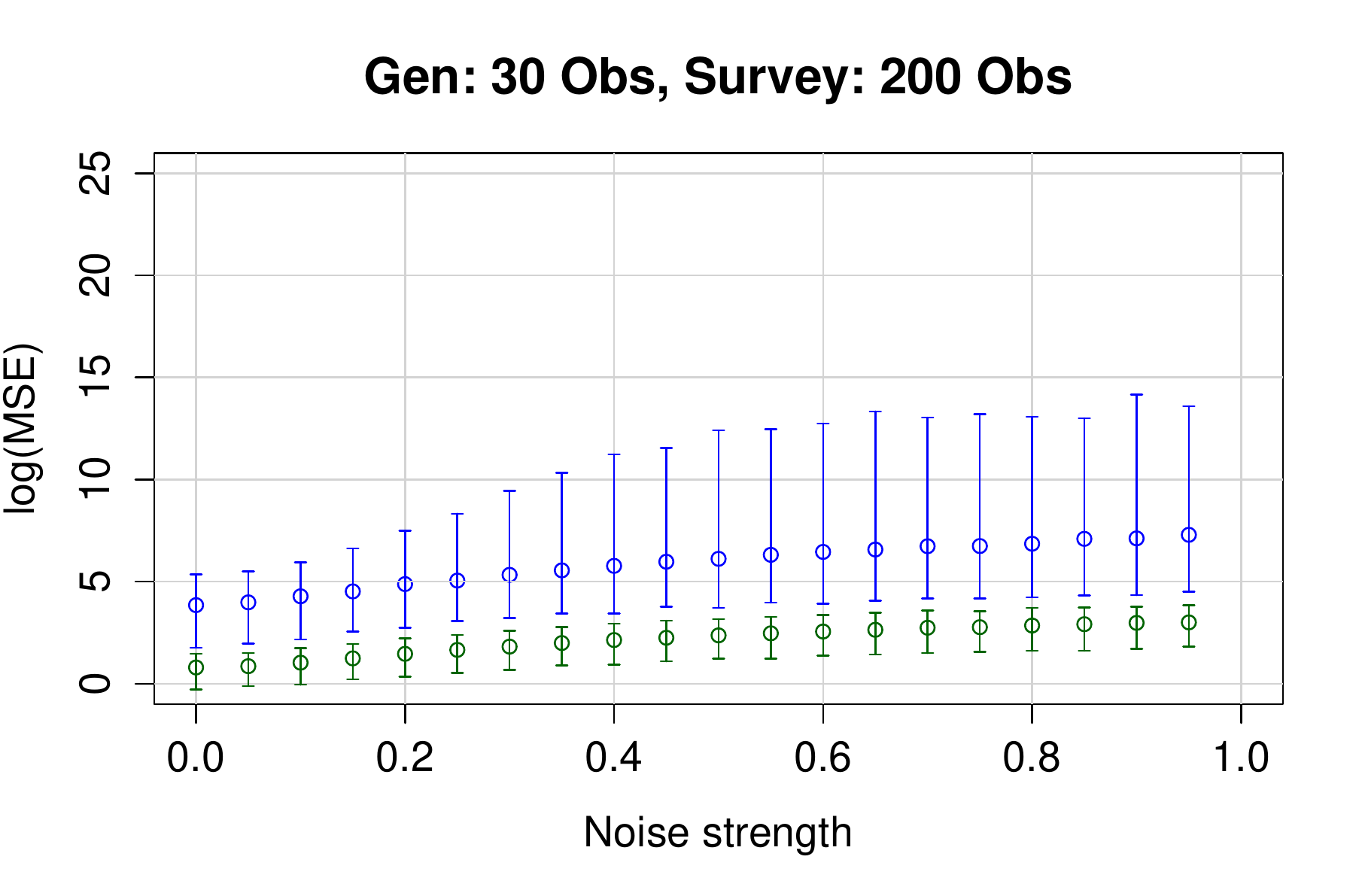}
\includegraphics[width=7cm, height=6cm]{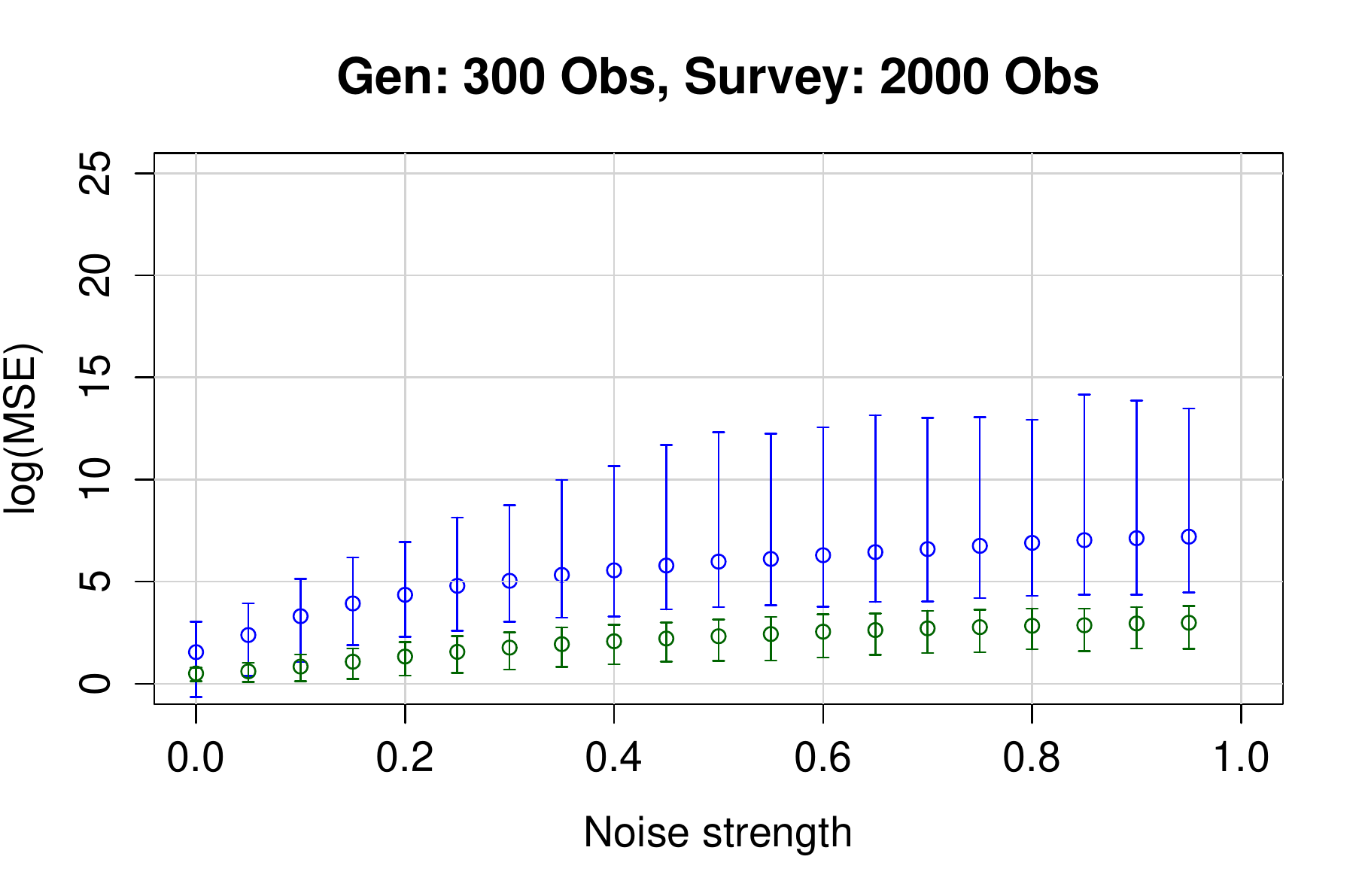}
\caption{\small Robustness of the estimators to the accuracy of the survey through the estimators' MSE. The eigen-value based estimation is plotted blue with $95 \%$ CI. The Ad-Hoc estimation is plotted in green. The upper line stands for a bias designed according to Eq. \ref{eq:noise}, and the bottom line for Eq. \ref{eq:prop.noise}. The MSE is calculated among $2000$ simulations. }
\label{fig:SensNoise}
\end{figure}

We define the robustness of our estimation as its ability to overcome biased survey results. To analyse the performance of the estimators, we considered that the parameter matrix of the survey was biased according to the following two equations,
\begin{eqnarray}
\label{eq:noise}
R_{survey} = \alpha R_z + \eta \times \mathcal{P}(\max(\alpha R_z))
\end{eqnarray}
where $\alpha \in [0,1]$ stands for the scale of the survey, $\eta \in [0,1]$ stands for the varying noise level, and $\mathcal{P}$ means the Poisson distribution, and,
\begin{eqnarray}
\label{eq:prop.noise}
R_{survey} = \alpha R_z + \eta \times \mathcal{P}(\alpha R_z)
\end{eqnarray}
The difference between the two equations relies in the bias structure. In Eq. \ref{eq:noise}, integer values are randomly added regardless of the real value of the parameter. It is the kind of errors we expect to find in badly designed surveys or respondents responding randomly. In Eq. \ref{eq:prop.noise}, we consider that the bias keeps the same structure as the $R_z$ matrix. This is more an equivalent of a measurement error usually described in the literature with centred Gaussian distributions. Therefore, we expect the estimators to provide better estimates when the survey's parameters are driven by Eq. \ref{eq:prop.noise}.\\
Figure \ref{fig:SensNoise} shows the ML and Ad Hoc estimation results for the two kind of bias, with different number of observations and different strength of bias. As expected, the second type of error (eq. \ref{eq:prop.noise}) is more easily overcome by the estimators. The important conclusion we can make from this result is that the Ad Hoc estimation performs better than the naive ML estimation, no matter what kind of noise we add.\\
Finally, and also very important, the Ad Hoc estimator seems more consistent and its performances are less affected by the number of available observations. For instance, we can observe that the log(MSE) profile is identical in Fig. \ref{fig:SensNoise} for $30-200$ observations and for $300-2000$. The importance of that being to allow some reliable time-dependency analysis (on a monthly basis, it would mean $30$ observations), which would be more complicated with the ML estimation.  

\section{Application}
\label{sec:app}

\subsection{Simulation Study}

Let M be a $5 \times 5$ matrix, representing departure and arrival stations (origin and destination). We assume that any value of M is a random number, generated according to the matrix of parameters $\mathcal{M}$, and the probability $p$. The matrix $\mathcal{M}$ has the following values,
\begin{equation}
\left[
\begin{array}{ccccc}
0 & 70 & 11 & 54 & 51 \\
70 & 0 & 23 & 43 & 82 \\
11 & 23 & 0 & 95 & 13 \\
54 & 43 & 95 & 0 & 22 \\
51 & 82 & 13 & 22 & 0
\end{array}
\right]
\end{equation}
Let $\tilde{M}$ be an observed sub-sample of M, i.e. only a proportion $\pi$ of M is represented in $\tilde{M}$. We assume that for every value of M, $\pi$ will be the same, only perturbed by some low-level additive noise. A realization of $\tilde{M}$ is the following matrix,
\begin{equation}
\left[
\begin{array}{ccccc}
0 & 11 & 3 & 10 & 7 \\
7 & 0 & 5 & 8 & 14 \\
4 & 2 & 0 & 14 & 3 \\
5 & 8 & 18 & 0 & 2 \\
3 & 11 & 8 & 8 & 0
\end{array}
\right]
\end{equation}
The value of $\pi$ in this example is roughly equal to $1/6$. The estimation of $\tilde{\mathcal{M}}$ is performed with the observations of this matrix. Then, $\tilde{\mathcal{M}}$ being symmetric, we diagonalize it, and using the optimization program described in section \ref{sec:OI} we are able to provide the following estimated $\hat{\mathcal{M}}$ matrix (for $10$, $50$, $500$ and $10000$ observations respectively),
\begin{equation}
\begin{array}{cc}
\left[
\begin{array}{ccccc}
0 & 60 & 48 & 48 & 54 \\
60 & 0 & 18 & 42 & 287 \\
48 & 18 & 0 & 60 & 12 \\
48 & 42 & 60 & 0 & 90 \\
42 & 287 & 12 & 90 & 0
\end{array}
\right]
&
\left[
\begin{array}{ccccc}
0 & 68 & 24 & 42 & 54 \\
68 & 0 & 18 & 36 & 124 \\
24 & 18 & 0 & 126 & 12 \\
42 & 36 & 126 & 0 & 18 \\
54 & 124 & 12 & 18 & 0
\end{array}
\right]
\\
& 
\\
\left[
\begin{array}{ccccc}
0 & 78 & 12 & 60 & 42 \\
78 & 0 & 24 & 48 & 78 \\
12 & 24 & 0 & 102 & 12 \\
60 & 48 & 102 & 0 & 24 \\
42 & 78 & 12 & 24 & 0
\end{array}
\right]
&
\left[
\begin{array}{ccccc}
0 & 72 & 12 & 54 & 48 \\
72 & 0 & 24 & 42 & 78 \\
12 & 24 & 0 & 96 & 12 \\
54 & 42 & 96 & 0 & 24 \\
48 & 78 & 12 & 24 & 0
\end{array}
\right]
\end{array}
\end{equation}
Table \ref{tble:MSE_bin} provides the mean square error of the estimator for different number of observations and different value of dispersion in the case of a negative binomial modelling. We also display in table \ref{tble:norm_test} the p-values for the Cramer-von Mises normality test of $1000$ estimation procedures for each eigenvalue.
\begin{table}[!ht]
\normalsize
\centering
\begin{tabular}{c|cccc}
  \hline
   Num. Obs. &  10 & 50 & 500 & 10000\\
  \hline
  \hline
    p = 0.2	 & $1575$ {\small ($>7\times10^6$)}& $101.1$ {\small ($7538$)}& $11.13$ {\small ($43.39$)}& $3.008$ {\small ($4.318$)}\\
  0.5   & $5888$ {\small ($>1\times10^7$)} & $441.9$ {\small ($>4\times10^5$)} & $24.90$ {\small ($302.4$)} & $3.112$ {\small ($3.084$)}\\
  0.8  & $4169$ {\small ($>2\times10^6$)} & $2972$ {\small ($>2\times10^6$)} & $193.3$ {\small ($>5\times10^4$)} & $9.743$ {\small ($94.70$)}\\
   \hline
\end{tabular}
\caption{\small \label{tble:MSE_bin} Negative Binomial modelling: Values of the mean square error of the parameter estimation with different number of observations and different distribution assumption over $1000$ replication of the simulation, and the according variance (under brackets).}
\end{table}
The table \ref{tble:MSE_pois} provides the mean square error of the estimator for different number of observations in the case of a Poisson modelling. 

The results of our estimation method on this simulation example seem very promising and we applied it to a real case data in the next paragraph.
 \begin{table}[!ht]
\normalsize
\centering
\begin{tabular}{c|cccc}
  \hline
  Num. Obs. & 10 & 50 & 500 & 10000\\
  \hline
  \hline
 MSE & $13.82$ {\small ($47.16$)} & $4.196$ {\small ($7.024$)} & $1.888$ {\small ($0.078$)} & $1.838$ {\small ($0.077$)}\\
  \hline
\end{tabular}
\caption{\small \label{tble:MSE_pois} Poisson modelling: Values of the mean square error of the parameter estimation with different number of observations over $1000$ replication of the simulation, and the according variance (under brackets).}
\end{table}
 \begin{table}[!ht]
\normalsize
\centering
\begin{tabular}{c|cccc}
  \hline
  Num. Obs. & 10 & 50 & 500 & 10000\\
  \hline
  \hline
 $\lambda_1$ & $0.071$ {\small ($0.27$)} & $0.034$ {\small ($0.78$)} & $0.031$ {\small ($0.82$)} & $0.042$ {\small ($0.65$)}\\
 $\lambda_2$ & $3.291$ {\small ($10^{-10}$)} & $1.203$ {\small ($10^{-10}$)} & $0.089$ {\small ($0.16$)} & $0.114$ {\small ($0.07$)}\\
 $\lambda_3$ & $5.291$ {\small ($10^{-10}$)} & $1.413$ {\small ($10^{-10}$)} & $0.284$ {\small ($0.001$)} & $0.103$ {\small ($0.11$)}\\
 $\lambda_4$ & $0.283$ {\small ($10^{-4}$)} & $0.050$ {\small ($0.51$)} & $0.053$ {\small ($0.47$)} & $0.079$ {\small ($0.21$)}\\
 $\lambda_5$ & $0.092$ {\small ($0.14$)} & $0.104$ {\small ($0.10$)} & $0.055$ {\small ($0.44$)} & $0.028$ {\small ($0.87$)}\\
  \hline
\end{tabular}
\caption{\small \label{tble:norm_test} Asymptotic behaviour of the estimator: Values of Cramer-von Mises Statistic for $1000$ replication of the simulation, and the corresponding p-value (in brackets).}
\end{table}
\\
Table \ref{tble:MSE_Reg} displays the mean square error of the estimators for the regression modelling. As we can see, the estimator is consistent. And we also can conclude that the estimator seems to behave reasonably well for $500$ or more observations. This number will however depend on the size of the exogenous parameter space and explanatory "power" in terms of $\%$ variation explained.
 \begin{table}[!t]
\normalsize
\centering
\begin{tabular}{c|cccc}
  \hline
   Num. Obs. & 10 & 50 & 500 & 10000\\
  \hline
  \hline
MSE${}_{\beta_0}$ & $8.10^7$ {\small ($6.10^{22}$)} & $1.10^4$ {\small ($1.10^{12}$)} & $3.10^3$ {\small ($8.10^7$)} & $5.10^2$ {\small ($2.10^5$)}\\
MSE${}_{\beta_1}$ & $7.10^6$ {\small ($4.10^{16}$)} & $6.10^3$ {\small ($3.10^7$)} & $5.10^2$ {\small ($2.10^5$)} & $2.10^1$ {\small ($6.10^2$)}\\
MSE${}_{\beta_2}$ & $9.10^7$ {\small ($9.10^{18}$)} & $7.10^3$ {\small ($5.10^7$)} & $6.10^2$ {\small ($3.10^5$)} & $3.10^1$ {\small ($6.10^2$)}\\
 \hline
\end{tabular}
\caption{\label{tble:MSE_Reg} Regression modelling: Values of the mean square error of the regression estimation ($M = 3$) with different number of observations over $1000$ replication of the simulation, and the according variance (in brackets).}
\end{table}
\subsection{Sydney Region Ferry}

\begin{figure}[htb]
\centering
\includegraphics[trim=0.0cm 2cm 1.2cm 0.0cm, clip=true,width=6cm, height=6cm]{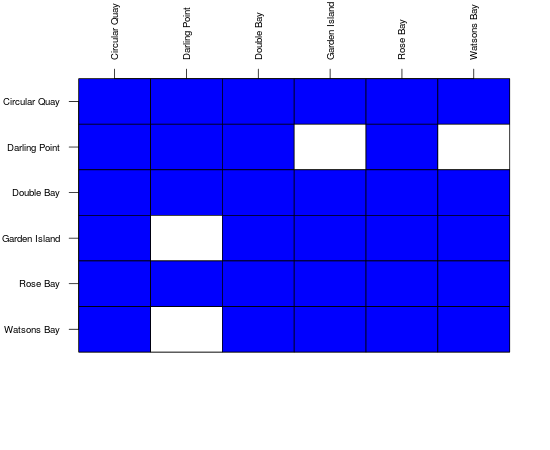}
\includegraphics[trim=0.0cm 2cm 1.2cm 0.0cm, clip=true, width=6cm, height=6cm]{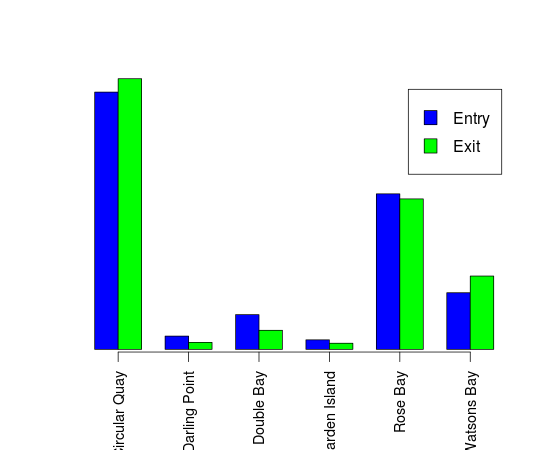}
\caption{\small Ferry Links and Passengers Entry/Exit by wharf in 2010}
\label{fig:LPW20xx}
\end{figure}

\begin{figure}[htb]
\centering
\includegraphics[width=6.5cm, height=6.5cm]{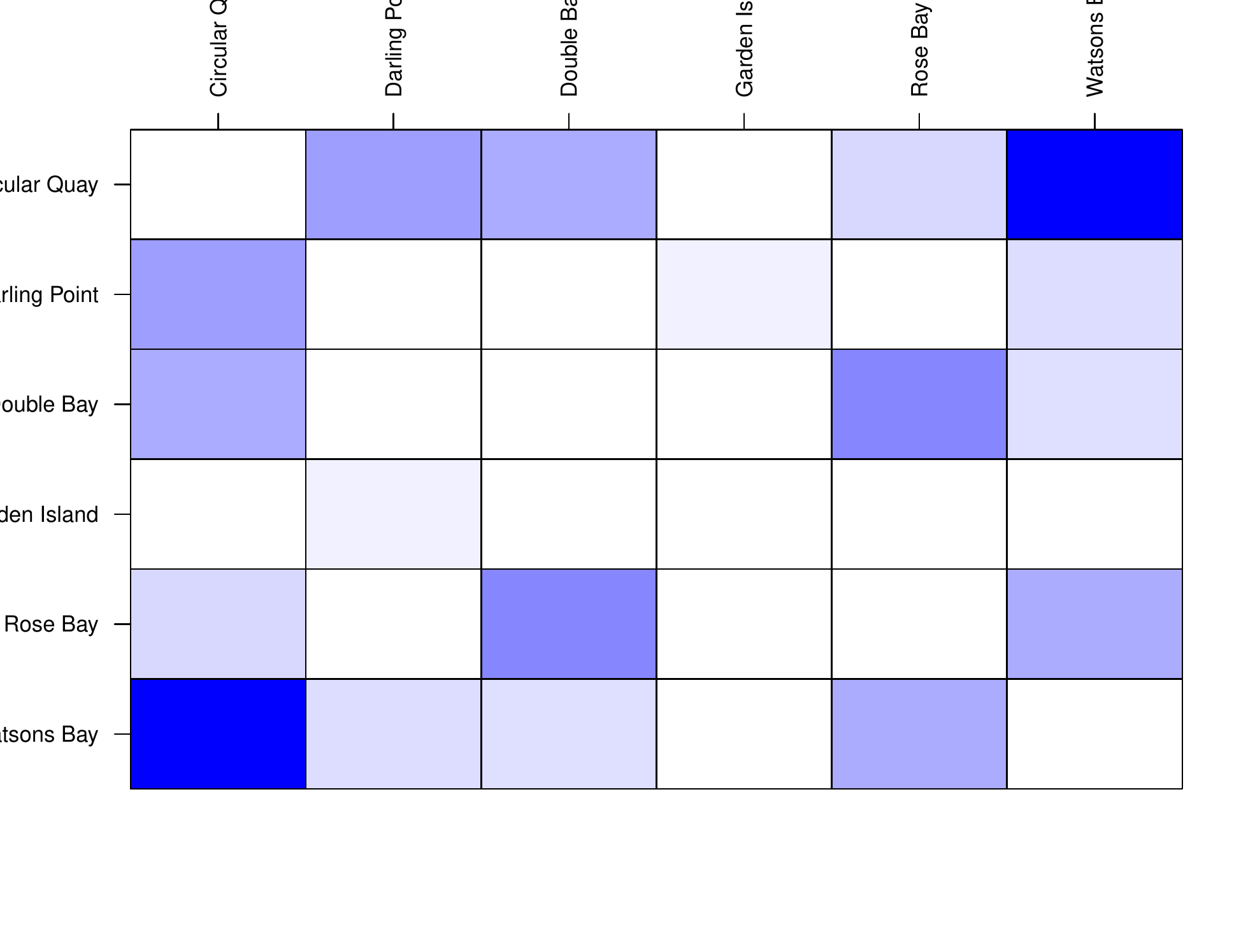}
\includegraphics[width=6.5cm, height=6.5cm]{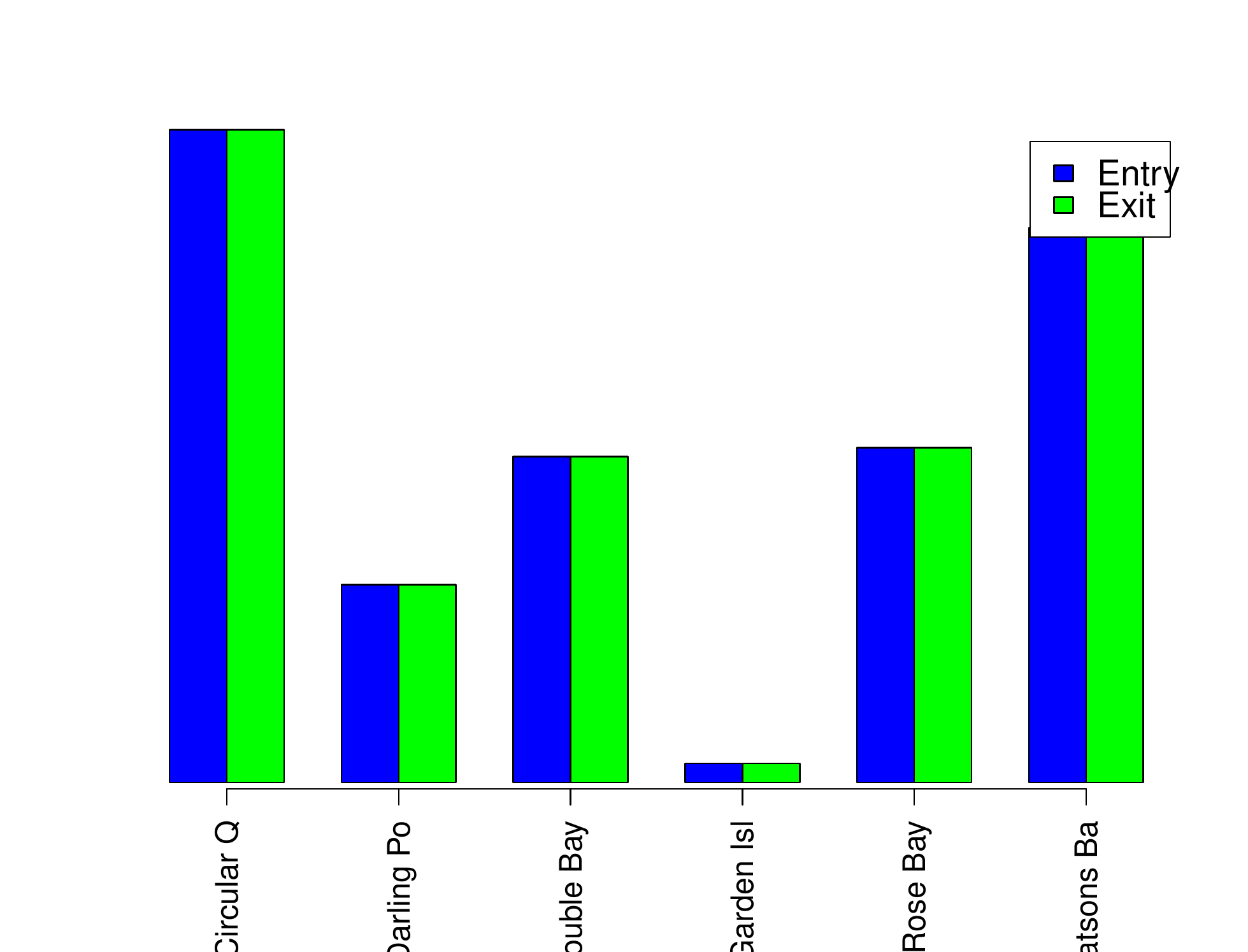}
\caption{\small OD Expectation matrix and Expected Passengers Entry/Exit after the estimations of the parameters, according to the Ad Hoc estimation.}
\label{fig:est2}
\end{figure}

We consider as an example the passengers commuting from the Sydney Region Ferry, according to five surveys organized in $2010$, $2011$, and $2012$ by The Bureau of Transport Statistics of New South Wales. The overall survey could have been analysed, but in order to be understandable, we will focus our analysis on the Eastern Suburbs route, composed with $6$ different wharfs and $35$ different days. Fig. \ref{fig:LPW20xx} represents the links between the wharfs, meaning that you can have a direct access by taking only one ferry.\\
As we can see from the plots, all the wharfs seem to be able to be reached starting from any other wharf, except for Darling Point, Watsons Bay and Garden Island. \\
Starting from this, the objective is the estimation of the OD matrix, which is almost like the Ferry Links matrix presented in the figures except that instead of logical values, it will be filled by counts of passengers. However, this survey does not provide enough data to perform this estimation. In particular, it doesn't give any observation on the ticket sales at the Wharfs. Then, to provide an estimation, we use a different survey from the same institute, which aimed at understanding the preferred mean of transport for people going to work. From this database, we only kept those who declared taking the ferry.\\
\begin{figure}[htb]
\centering
\includegraphics[trim=0.0cm 2cm 1.2cm 0.0cm, clip=true,width=6cm, height=6cm]{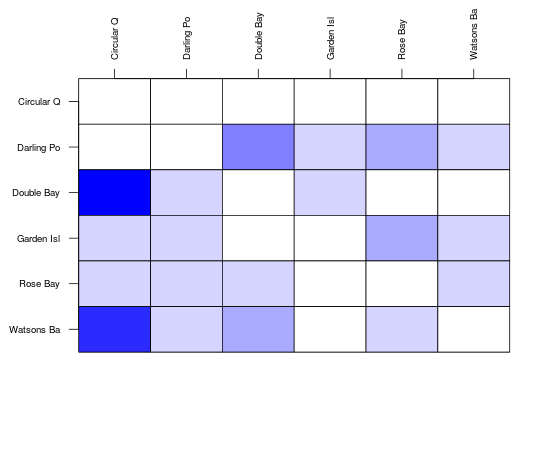}
\includegraphics[trim=0.0cm 2cm 1.2cm 0.0cm, clip=true,width=6cm, height=6cm]{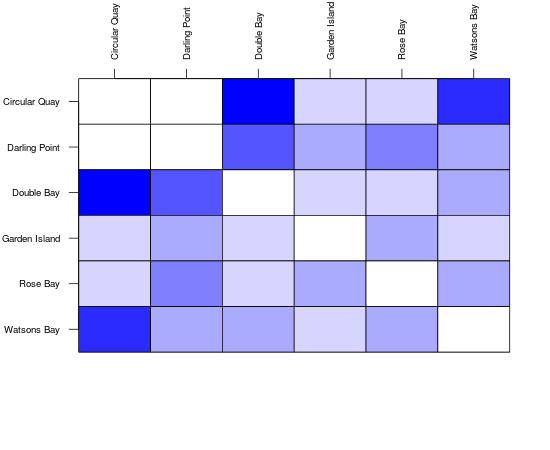}
\caption{\small Left: Observations OD matrix of the Survey. Right: Estimated OD parameter matrix of the survey. The darker, the greater number of passengers.}
\label{fig:JW2006}
\end{figure}
Before we could provide an analysis of these data, we had to consider that the origins and destination of the people in the second survey didn't correspond to the wharfs of the first survey. Therefore, we decided to attribute to every passenger the starting point of their journey as the closest wharf to their home, and a destination point the closest wharf to their office. The distance have been calculated according to each Wharf and Location longitude and latitude values.
This being done, we can provide a new origin-destination matrix. It is plotted in Fig. \ref{fig:JW2006}.\\
\begin{figure}[htb]
\centering
\includegraphics[trim=0.0cm 2cm 1.2cm 0.0cm, clip=true,width=6cm, height=5.5cm]{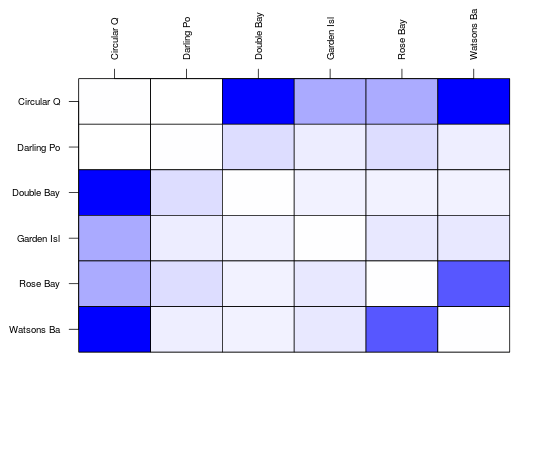}
\includegraphics[trim=0.0cm 2cm 1.2cm 0.0cm, clip=true, width=6cm, height=5.5cm]{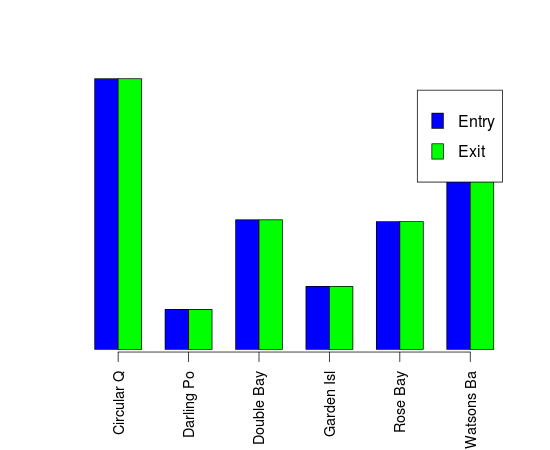}
\caption{\small OD Expectation matrix and Expected Passengers Entry/Exit after the estimations of the parameters, according to the naive ML estimation.}
\label{fig:est1}
\end{figure}
 This OD matrix is then supposed to have the same structure as the regular OD of the barrier counts. Therefore, we will apply the methodology presented in section \ref{sec:CPE} to reconstruct the 2010, 2011 and 2012 OD matrix according to the barrier counts. The eigenvalues and eigenvectors are then calculated, and the reconstruction of the Ferry passengers are presented in Figs. \ref{fig:est1} and \ref{fig:est2}. 

\section{Conclusion and further work}

We presented in this paper a new estimation technique for the OD matrix. We use the information available from surveys to infer the correct projection matrix and reduce significantly the size of the parameter space. By using a maximum likelihood approach, we compute the estimating equations and provide an Ad Hoc estimator of the OD matrix. We demonstrated its robustness in section \ref{sec:rob}.\\
We also demonstrated that a regression analysis could be performed on this kind of data, and showed that this estimation procedure is also consistent. To the best of our knowledge, this is the first time that a such multivariate approach is used to estimate the OD matrix. This approach will improve the prediction ability of passengers journeys.\\
We finally applied our techniques to simulated data and real-case scenario in Sydney ferry transport using the data from the Bureau of Transport Statistics.\\
The estimation of the OD matrix is a first step for the analysis of the passengers flow over the transport network. Then, beyond this estimation point, we may cite:
\begin{enumerate}
\item \label{monit} Monitoring the passengers count;
\item \label{forecast} Forecasting of the passengers count (1 week in advance for example);
\item \label{predict} Predicting the passengers flow in case of spatio-temporal topological change in the network
\end{enumerate}

In order to address the monitoring task (\ref{monit}), several strong assumptions have to be made, which will require ground verifications before being tested. Among them, we can cite the time between ticket validation and getting into the train or the regularity at which people take their train if they are regular passengers. Moreover, a real-time access to the data is necessary. While difficult, this seems achievable.\\
The forecasting of passenger counts (\ref{forecast}) can be done without additional information (if sufficient temporal information has been provided in a first place), even if more observation would probably mean smaller variances. These forecasts could be helpful for efficient scheduling the trains (for example), but further study have to be done in order to understand the influence of complex variables such as the temperature or the humidity.\\
Real-time prediction of passenger flows (\ref{predict}) is more difficult, but is theoretically achievable. What we denote a spatio-temporal topological change is a change in the timetable, or in the public transport route.

\bibliographystyle{model2-names}


\newpage
\appendix

\section{}
\label{ann:A}

{ \small \paragraph{\bf Calculation in case of Poisson distribution}
\emph{To make it clear, we will explore a simple density example, where the Poisson distribution will be used instead of the Negative Binomial. Therefore, the pdf can be expressed as,
\begin{eqnarray}
f_{R_s}(\mathcal{X}_R) &=& \frac{\mu^{\mathcal{X}_R}}{\mathcal{X}_R!} e^{-\mu}
\end{eqnarray}
where $\mu$ is the parameter we are interested in. Then, if we make the assumption that the number of passengers in every station are independent, each $\tilde{y}_D^i$ is distributed according to a Poisson distribution, with the parameter $\sum_j r_{ij}$, which can be re-written according to Eq. \ref{eq:diag2},
\begin{eqnarray}
\tilde{y}_D^i & \sim & \mathcal{P}(\sum_k \lambda_k p_{ik} \sum_j p_{jk})
\end{eqnarray}
where the $\lambda_i$ are the eigenvalues, and the $p_{ij}$ the element of the matrix P. If we denote $s_k = \sum_j p_{jk}$, the transformed density can then be written,
\begin{eqnarray}
g_{R_z}(\tilde{y}_D) &=& \frac{(\sum_k \lambda_k p_{ik} s_k)^{\tilde{y}_D}}{\tilde{y}_D!} e^{-\sum_k \lambda_k p_{ik} s_k}
\end{eqnarray}
\\
Therefore, the log-likelihood can be expressed as follow,
\begin{eqnarray}
log \mathcal{L} & = & -N \sum_{k=1}^n \lambda_k s_k^2 + \sum_{i=1}^n \ln\big(\sum_{k=1}^n \lambda_k p_{ik} s_k\big) \sum_{l=1}^N \tilde{y}_i^l
\end{eqnarray}
The maximum likelihood estimation is then equivalent to solve the following system,
\begin{equation}
\label{eq:sys_pois}
\left\{
\begin{array}{cll}
-N s_1^2 + \sum_{i=1}^n \frac{p_{i1} s_1}{\sum_{k=1}^n \lambda_k p_{ik} s_k} \sum_{l=1}^N \tilde{y}_i^l &=& 0\\
 \vdots &\vdots & \\
-N s_n^2 + \sum_{i=1}^n \frac{p_{in} s_n}{\sum_{k=1}^n \lambda_k p_{ik} s_k} \sum_{l=1}^N \tilde{y}_i^l &=& 0
\end{array}
\right.
\end{equation}
still under the constraint {\bf C1} and {\bf C2}. {\bf C3} is excluded because this set of parameter doesn't exist in the Poisson modelling. If $n = 1$, ${\bf C1} \Leftrightarrow  \lambda \geq 0$ and {\bf C2} doesn't stand. Then estimated value corresponds to the classical one dimensional Poisson unbiased mean estimator $\hat{\lambda} = \bar{y}$.\\
The system of equations \ref{eq:sys_pois} seems at first a quite complicated one. Nevertheless, it can be simplify so as to become,
\begin{eqnarray}
\label{eq:sys_pois2}
\forall i, \quad -s_i + \sum_{j=1}^n \frac{p_{ji} \bar{y}_j}{\mu_j} &=& 0
\end{eqnarray}
where $\mu_j = \sum_k \lambda_k p_{jk} s_k$ contains the unknown parameters. Then, if we denote $U = (1/\mu_j)_j$, then,
\begin{eqnarray}
\label{eq:linsol_mu}
\hat{U} & = & \big( \bar{Y}_d P {}^t P \bar{Y}_d \big)^{-1} \bar{Y} P S
\end{eqnarray}
where $\bar{Y}_d = diag(\bar{y}_i)_{i}$ and $S = (s_i)_i$. Then, we can keep simplifying the expression,
\begin{eqnarray}
\label{eq:linsol_mu2}
\hat{U} & = & \bar{Y}_d^{-1} P S
\end{eqnarray}
Finally, the same reasoning leads to the following estimator,
\begin{eqnarray}
\label{eq:linsol_lam}
\hat{\Lambda} & = & \big( S_d {}^t P P S_d \big)^{-1} S_d {}^t P \Big(\bar{Y}_d^{-1} P S \Big)^{-1}
\end{eqnarray} 
where $S_d = diag(s_i)_{i}$. This estimator will probably not be the best estimator given that it relies on the inversion of $\hat{U}$, but has the advantage to be asymptotically unbiased, with variance decreasing to zero.}

{ \small \paragraph{ Properties of the calculated estimators}
\emph{Let $f$ be a probability density function. If $f_{\Lambda}$ denotes the pdf of $\Lambda$, and $f_P$ the pdf of $\tilde{P}$, we can write,
\begin{eqnarray}
f_{\Lambda} &=& \int f_{\Lambda \vert \tilde{P} }(\lambda) f_{\tilde{P}}(\tilde{p}) d\tilde{p}
\end{eqnarray}
Let consider the estimator presented in Eq. \ref{eq:linsol_lamfg}, and make the assumption that we are in a large value case, meaning $\mathcal{P}(.) \sim  \mathcal{N}(.,.)$. Then, 
\begin{eqnarray}
\Lambda \vert \tilde{P} & \sim & \mathcal{N}\big(m, N^{-1} \Sigma \big)
\end{eqnarray}
where,
\begin{equation}
\left\{
\begin{array}{lll}
m &=& \tilde{S}_d^{-1} {}^t\tilde{P} P S_d \Lambda \\
\Sigma &=& \tilde{S}_d^{-1} \big[ D + {}^t \tilde{P} diag(PS_d \Lambda) \tilde{P} \big] \tilde{S}_d^{-1}
\end{array}
\right.
\end{equation}
and $\tilde{P}$ is an estimation of $P$ according to the first observations.\\
~\\
}}}

\section{}
\label{ann:B}

To prove the convergence in probability, we need to demonstrate that,
\begin{eqnarray}
\lim_n \mathbb{P}(\vert \hat{\Lambda} - \Lambda \vert \geq \epsilon) & = & 0
\end{eqnarray}
where $n = \min(N_1,N_2)$. Starting with the left hand side, we have,
\begin{eqnarray}
\mathbb{P}(\vert \hat{\Lambda} - \Lambda \vert \geq \epsilon) & = & \mathbb{P}(\vert \tilde{S_d}^{-1} {}^t \tilde{P} \bar{Y} - \Lambda \vert \geq \epsilon)\nonumber\\
 & \leq & \mathbb{P}(\vert \tilde{S_d}^{-1} {}^t \tilde{P} \bar{Y} - S_d^{-1} {}^t P \bar{Y} \vert + \vert S_d^{-1} {}^t P \bar{Y} - \Lambda \vert \geq \epsilon)\nonumber\\
 &=& \int \mathbb{P}\Big(\vert S_d^{-1} {}^t P \bar{Y} - \Lambda \vert \geq \epsilon - c \Big)\mathbb{P}(\vert (\tilde{S_d}^{-1} {}^t \tilde{P} - S_d^{-1} {}^t P)\bar{Y} \vert = c) dc
\end{eqnarray}
And we know that $\bar{Y} \xrightarrow[]{\ a.s.\ } P S_d \Lambda $. Then,
\begin{eqnarray}
\label{eq:as}
\forall \epsilon > 0, \quad \lim_n \mathbb{P}\Big(\vert S_d^{-1} {}^t P \bar{Y} - \Lambda \vert \geq \epsilon \Big) &=& 0
\end{eqnarray}
and we have,
\begin{eqnarray}
\mathbb{P}(\vert \hat{\Lambda} - \Lambda \vert \geq \epsilon) &\leq & \int_0^\epsilon  \mathbb{P}\Big(\vert S_d^{-1} {}^t P \bar{Y} - \Lambda \vert \geq \epsilon - c \Big) f_n (c)dc + \int_\epsilon^{+\infty} f_n (c)dc
\end{eqnarray}
where $f_n()$ stands for the probability density function of $\vert (\tilde{S_d}^{-1} {}^t \tilde{P} - S_d^{-1} {}^t P)\bar{Y} \vert$. \\
The first integral decreases towards $0$ as $n$ grows to infinity according to Eq. \ref{eq:as}. The argument for the second integral is the following. According to the assumption of strong convergence of $\tilde{P}$, $f_n()$ converge towards the Dirac function $\delta_0()$ as $N_1$ goes to infinity. $\epsilon$ being strictly positive, this ends the proof. $\blacksquare$.

\section{}
\label{ann:C}

{ \small \paragraph{\bf Calculation in case of Poisson regression (and log link function)}
\emph{The beginning of the reasoning is similar to the previous one. Then, if we assume that exogenous variables have impacts on the number of passengers, we can write,
\begin{eqnarray}
R &=& {\bf \beta}_0 + \sum_m {\bf \beta}_m x_m
\end{eqnarray}
where $\beta$ are symmetric matrices reflecting the intercept ($\beta_0$) for baseline commuter flows and the variable influences ($\beta_m$) for changes in commuter flows from known daily influences. Moreover, we assume that the same diagonalization (meaning with the same eigenvectors) can be applied, which lead us to,
\begin{eqnarray}
r_{ij} &=& \underbrace{\sum_k d_k^0 p_{jk} p_{ik}}_{\textrm{fixed part}} + \underbrace{\sum_k \sum_m d_k^m p_{jk} p_{ik} x_m}_{\textrm{multivariate-time varying part}}
\end{eqnarray} 
Therefore, $\tilde{y}_i$ will be distributed according to a Poisson distribution with the following parameter,
\begin{eqnarray}
\sum_j r_{ij} &=& \sum_k d_k^0 p_{ik} s_k + \sum_k \sum_m d_k^m p_{ik} s_k x_m
\end{eqnarray}
where the parameters to be estimated are $(d_k^0)_k,(d_k^m)_{k,m}$, which means we have to estimate $n \times (r+1)$ parameters.\\
The probability of one observation can then be written,
\begin{eqnarray}
p(\tilde{y}_i^l \vert x_m^l) &=& \frac{ ( \sum_k d_k^0 p_{ik} s_k + \sum_k \sum_m d_k^m p_{ik} s_k x_m^l)^{\tilde{y}_i^l}}{\tilde{y}_i^l ! } \times \nonumber \\
&&  e^{-(\sum_k d_k^0 p_{ik} s_k + \sum_k \sum_m d_k^m p_{ik} s_k x_m^l)}
\end{eqnarray}
which gives the following log-likelihood,
\begin{eqnarray}
log \mathcal{L} & \propto & \sum_i \Bigg(\sum_l \tilde{y}_i^l \log\Big(\sum_k d_k^0 p_{ik} s_k + \sum_k \sum_m d_k^m p_{ik} s_k x_m^l \Big) \nonumber\\
&& - \sum_l \Big( \sum_k d_k^0 p_{ik} s_k + \sum_k \sum_m d_k^m p_{ik} s_k x_m^l \Big) \Bigg)
\end{eqnarray}
Therefore, to obtain the final system of equation, we need to calculate the derivatives of the log-likelihood with respect to each parameter $d_k^m$.
}}

\end{document}